\documentclass[aps,12pt,final,notitlepage,onecolumn]{revtex4}

\usepackage{epsfig}
\usepackage{amsmath}
\usepackage{hyperref}

\begin{document}

\title{Comparison of $\psi(2S)$ and $J/\psi$ photoproduction cross sections 
in Pb-Pb ultraperipheral collisions at the LHC}

\author{\firstname{Vadim} \surname{Guzey}}
\email{vguzey@pnpi.spb.ru}
\affiliation{Petersburg Nuclear Physics Institute (PNPI), \\
National Research Center ``Kurchatov Institute'', Gatchina, 188300, Russia}

\author{\firstname{Michael} \surname{Zhalov}}
\email{zhalov@pnpi.spb.ru}
\affiliation{Petersburg Nuclear Physics Institute (PNPI), \\
National Research Center ``Kurchatov Institute'', Gatchina, 188300, Russia}

\begin{abstract}

Within the leading logarithmic approximation of perturbative QCD and the leading twist
approach to nuclear shadowing,
we compare the cross sections of photoproduction of $J/\psi$ and $\psi(2S)$ mesons on nuclei in
Pb-Pb UPCs in the LHC kinematics.
We find that the nuclear suppression of these cross sections
due to the nuclear gluon shadowing is large and rather similar 
in the $J/\psi$ and $\psi(2S)$ cases.

\end{abstract}

\maketitle

In a recent series of papers~\cite{Guzey:2013xba,Guzey:2013qza} it was argued that the ALICE data
on $J/\psi$ photoproduction in Pb-Pb ultraperipheral collisions (UPCs) at the LHC at $\sqrt{s}=2.76$ 
GeV~\cite{Abbas:2013oua,Abelev:2012ba}
constrains the nuclear gluon distribution at small $x$ and favors the significant nuclear gluon shadowing
at $x \approx 10^{-3}$. This note presents an extension of the analysis of~\cite{Guzey:2013qza} to
the case of photoproduction of $\psi(2S)$ vector meson, the first excitation of $J/\psi$,   
in Pb-Pb UPCs at the LHC. We show that the nuclear suppression is rather similar in 
the $\psi(2S)$ and  $J/\psi$ cases, which provides additional constrains on the dynamics and magnitude
of the nuclear gluon shadowing at $x \approx 10^{-3}$.

In QCD the cross sections of photoproduction of $J/\psi$ and $\psi(2S)$ vector mesons on 
the nucleon and nuclei can be treated on equal footing. 
In this work we use the leading logarithmic approximation of 
perturbative QCD, where the $\gamma+ T \to J/\psi [\psi(2S)] + T$
cross section is proportional to the gluon density of the target $G_T(x,\mu^2)$ squared~\cite{Ryskin:1992ui}:
\begin{equation}
\frac{d \sigma_{\gamma T \to J/\psi [\psi(2S)] T}}{dt}(W_{\gamma p},t_{\rm min})=C(\mu^2) [\alpha_s(\mu^2) x G_T(x,\mu^2)]^2
F_T^2(t_{\rm min}) \,,
\label{eq:cs}
\end{equation}
where $T$ denotes the nucleon or a nucleus; $W_{\gamma p}$ is the invariant photon--target center 
of mass energy
per nucleon; $t$ is the four-momentum transfer squared
and $t_{\rm min}=-x^2 m_N^2$ is its minimal kinematically-allowed value, where $m_N$ is the nucleon mass; 
$\mu$ is the factorization scale; $x=M_V^2/W_{\gamma p}^2$ is the light-cone momentum fraction 
associated with the two-gluon ladder, where
$M_V$ is the vector meson mass $(V=J/\psi$ or $\psi(2S)$);
$C(\mu^2)$ determines the cross section normalization and depends 
on the wave function of the final charmonium state and approximations used in the calculation of the 
strong $\gamma+ T \to J/\psi [\psi(2S)] + T$ amplitude;
$F_T(t)$ is the form factor of the target.
While for the proton target $F_p(t)=1$ with the very good accuracy, the effect of $t_{\rm min} \neq 0$ 
is important
for heavy nuclei. 

In the case of $J/\psi$ photoproduction on the proton, 
Eq.~(\ref{eq:cs}) was first derived in \cite{Ryskin:1992ui} using the non-relativistic approximation 
for the $J/\psi$ wave function and later extended beyond the non-relativistic and collinear
approximations in~\cite{Ryskin:1995hz,Martin:2007sb}. It was found that
$\mu^2 ={\cal O} (m_c^2)$, where $m_c$ is the charm quark mass
($\mu^2 = M_{J/\psi}^2/4=2.4$ GeV$^2$ in the non-relativistic limit) and 
$C(\mu^2)=F^2(\mu^2)(1+\eta^2) R_g^2 \pi^3 \Gamma_{ee} M_{J/\psi}^3/(48 \alpha_{\rm em} \mu^8)$, 
where $\Gamma_{ee}$ is the $J/\psi \to e^{+} e^{-}$ decay width; 
$\alpha_{\rm em}$ is the fine-structure constant; $\eta$ is the ratio of the real to the imaginary parts 
of the $\gamma + p \to J/\psi+p$ scattering amplitude; $R_g$ is the skewness factor taking into account 
the off-forward nature of the $\gamma + p \to J/\psi+p$ amplitude.
The factor of $F^2(\mu^2) \leq 1$ takes into account the relativistic corrections in the 
$J/\psi$ wave function and next-to-leading order corrections in $\alpha_s$.

Using the freedom to vary the factorization scale $\mu$ within a reasonable range, $2.4 \leq \mu^2 \leq 3.4$
GeV$^2$, it was shown in~\cite{Guzey:2013qza} 
that Eq.~(\ref{eq:cs}) with $\mu^2 \approx 3$ GeV$^2$ reproduces correctly the 
$W_{\gamma p}$ dependence of the 
$\gamma+ p \to J/\psi + p$ cross section measured at HERA and LHCb with a wide array of modern 
gluon distributions of the proton.
This emphasizes that high-energy diffractive photoproduction of charmonium 
places an important constraint on the small-$x$ behavior of $G_p(x,Q^2)$ down to $5 \times 10^{-6}$.

A similar analysis can be carried out in the case of $\psi(2S)$ photoproduction on the proton.
This is supported by the H1~\cite{Adloff:2002re} and ZEUS~\cite{Zulkaply:2012oaa}
results on diffractive photoproduction of $\psi(2S)$ 
mesons which found that the energy behavior of the diffractive $\psi(2S)$ 
cross section is similar to or possibly somewhat steeper than that for $J/\psi$ mesons:
\begin{equation}
R \equiv \frac{\sigma(\psi(2S))}{\sigma(J/\psi)} \propto \left(\frac{W_{\gamma p}}{90 \ {\rm GeV}}\right)^{\Delta  \delta} 
\label{eq:H1_2s}
\end{equation}
where $\Delta \delta=0.24 \pm 0.17$ on the interval $40 < W_{\gamma p} < 150$ GeV~\cite{Adloff:2002re}.
Using as a representative example the leading order CTEQ6L1 parton distribution~\cite{Pumplin:2002vw},
we find that Eq.~(\ref{eq:cs}) reproduces Eq.~(\ref{eq:H1_2s}) with $3.8 < \mu^2 < 4.5$ GeV$^2$, 
where the range of the $\mu^2$ values corresponds to the large experimental error on  $\Delta \delta$. 
In view of this uncertainty, in the analysis below, for definiteness, we use $\mu^2 = 4$  GeV$^2$ 
for $\psi(2S)$ photoproduction on the proton and nuclei.

Note that the exclusive $\psi(2S)$ photoproduction cross section within the leading and 
next-to-leading orders of perturbative QCD was analyzed in~\cite{Jones:2013eda} and predictions 
for $\psi(2S)$ photoproduction in proton--proton UPCs at the LHC were given.
Our results for the $W_{\gamma p}$ dependence of the cross section of diffractive 
$\psi(2S)$ photoproduction broadly agree with those of~\cite{Jones:2013eda}.

Applying Eq.~(\ref{eq:cs}) to the nucleus and proton targets and integrating the nucleus cross section
over the momentum transfer $t$, we obtain for the cross section of $J/\psi$ and $\psi(2S)$ 
photoproduction on a nucleus~\cite{Guzey:2013qza}:
\begin{equation}
\sigma_{\gamma A \to J/\psi [\psi(2S)] A}(W_{\gamma p})=\frac{C_A(\mu^2)}{C_p(\mu^2)} 
\frac{\sigma_{\gamma p \to J/\psi [\psi(2S)]p}}{dt}(W_{\gamma p},t=0) 
\left(\frac{x G_A(x,\mu^2)}{A x G_p(x,\mu^2)}\right)^2 \Phi_A(t_{\rm min}) \,,
\label{eq:cs2}
\end{equation}
where $\Phi_A(t_{\rm min})=A^2\int \limits_{t_{\rm min}}^{\infty} dt \left|F_A (t)\right|^2$ and
$F_A (t)$ is the nuclear form factor normalized to unity, $F_A (t=0)=1$. The suppression factor
$C_A(\mu^2)/C_p(\mu^2) \approx 1$, deviates from unity by a few percent and takes into account the fact
that nuclear shadowing slows down the $W_{\gamma p}$ dependence of the $\gamma+ A \to J/\psi [\psi(2S)] +A$
amplitude compared to that of the  $\gamma+ p \to J/\psi [\psi(2S)] +p$ one.

It is convenient to quantify the role of the nuclear gluon shadowing suppression in photoproduction of 
heavy vector mesons on nuclei by comparing predictions of Eq.~(\ref{eq:cs2}) with the nuclear cross section
calculated in the impulse approximation (IA): 
\begin{equation}
\sigma_{\gamma A \to J/\psi [\psi(2S)] A}^{\rm IA}(W_{\gamma p})= 
\frac{\sigma_{\gamma p \to J/\psi [\psi(2S)]p}}{dt}(W_{\gamma p},t=0) \Phi_A(t_{\rm min}) \,.
\label{eq:cs2_IA}
\end{equation}
Introducing the suppression factor $S(W_{\gamma p})$~\cite{Guzey:2013qza,Guzey:2013xba}, we obtain:
\begin{eqnarray}
S(W_{\gamma p}) \equiv \left[\frac{\sigma_{\gamma A \to J/\psi [\psi(2S)] A}(W_{\gamma p})}{\sigma_{\gamma A \to J/\psi [\psi(2S)] A}^{\rm IA}(W_{\gamma p})}\right]^{1/2} &=& \left[\frac{C_A(\mu^2)}{C_p(\mu^2)}\right]^{1/2} \frac{x G_A(x,\mu^2)}{A x G_p(x,\mu^2)} \nonumber\\
&=& \left[\frac{C_A(\mu^2)}{C_p(\mu^2)}\right]^{1/2} R(x,\mu^2) \,,
\label{eq:S}
\end{eqnarray}
where $R(x,\mu^2)=x G_A(x,\mu^2)/[A x G_p(x,\mu^2)]$ is the factor quantifying the nuclear gluon
shadowing.

\begin{figure}[h]
\centering
\epsfig{file=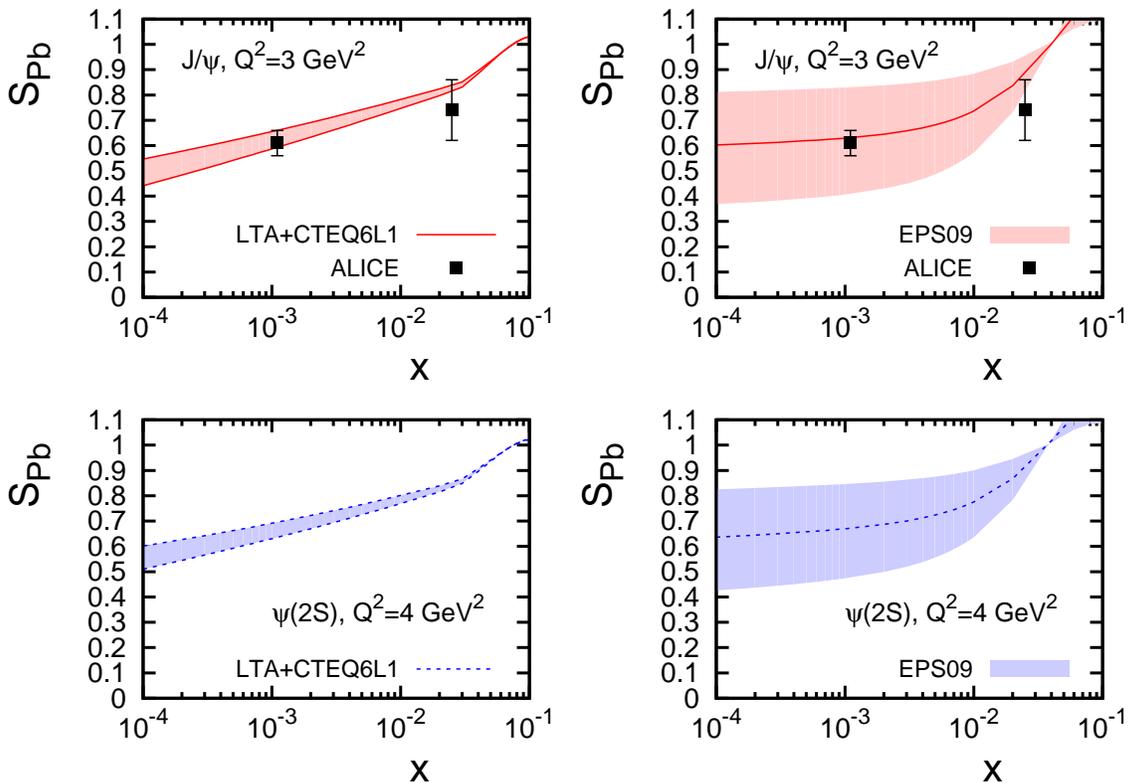,scale=1.5}
\caption{The suppression factor of $S(W_{\gamma p})$ of Eq.~(\ref{eq:S})
for photoproduction of $J/\psi$ (two upper panels) and $\psi(2S)$ (two lower panels) on $^{208}$Pb
as a function of $x=M_V^2/W_{\gamma p}^2$. 
We show two sets of theoretical predictions: those of the leading twist theory of nuclear 
shadowing~\cite{Frankfurt:2011cs} (the curves labeled ``LTA+CTEQ6L1'', which span  the theoretical
uncertainty band)
and those of the EPS09 global QCD fit of nuclear PDFs~\cite{eps09} (the central value 
and the associated shaded uncertainty band labeled ``EPS09'').
The filled squares and the associated errors are the results of the analysis of~\cite{Guzey:2013xba} 
in the $J/\psi$ case.
}
\label{fig:S_2s}
\end{figure}

Figure~\ref{fig:S_2s} presents our predictions for the suppression factor of $S(W_{\gamma p})$ 
for photoproduction of $J/\psi$ (two upper panels) and $\psi(2S)$ (two lower panels) on $^{208}$Pb
as a function of $x=M_V^2/W_{\gamma p}^2$. As we discussed above, the case of $J/\psi$ corresponds to 
$\mu^2=3$ GeV$^2$ and the case of $\psi(2S)$ corresponds to $\mu^2=4$ GeV$^2$. In the figure, 
we show two sets of predictions: the predictions of the dynamical leading twist theory of nuclear 
shadowing~\cite{Frankfurt:2011cs} (the curves labeled ``LTA+CTEQ6L1'', which span  the theoretical
uncertainty band)
and the results of the EPS09 global QCD fit of nuclear PDFs~\cite{eps09} (the central value 
and the associated shaded uncertainty band labeled ``EPS09'').

In the case of photoproduction of $J/\psi$, the theoretical predictions describe well 
the values of  $S(W_{\gamma p})$ (the filled squares with the associated errors), 
which were model-independently extracted in the analysis~\cite{Guzey:2013xba} 
of the ALICE data on $J/\psi$ photoproduction in Pb-Pb ultraperipheral collisions at the LHC at $\sqrt{s}=2.76$ TeV~\cite{Abbas:2013oua,Abelev:2012ba}.

One can see from Fig.~\ref{fig:S_2s} that the values of the suppression factor of $S(W_{\gamma p})$
for the  $J/\psi$ and $\psi(2S)$ cases are rather close. For instance, at the central rapidity $y=0$
of the vector meson produced in Pb-Pb UPCs at $\sqrt{s}=2.76$ TeV and the nucleus beam energy
$E_A=1.38$ TeV,
which correspond to
$x=M_{J/\psi}/(2 E_A)=1.1 \times 10^{-3}$ (for $J/\psi$) and $x=M_{\psi(2S)}/(2 E_A)=1.3 \times 10^{-3}$
(for $\psi(2S)$),
we obtain:
\begin{eqnarray}
&&S(x=1.1 \times 10^{-3}, \mu^2=3\ {\rm GeV}^2)=\left\{\begin{array}{ll}
0.59-0.66 \,, & {\rm  LTA+CTEQ6L1} \\
0.63 \pm 0.20 \,, & {\rm EPS09} \,, 
\end{array} \right. \nonumber\\
&&S(x=1.3 \times 10^{-3}, \mu^2=4\ {\rm GeV}^2)=\left\{\begin{array}{ll}
0.63-0.69 \,, & {\rm  LTA+CTEQ6L1} \\
0.67 \pm 0.18 \,, & {\rm EPS09} \,,
\end{array} \right.
\label{eq:S_y0}
\end{eqnarray} 
where the spread in the values of $S$ correspond to the theoretical uncertainty of
the LTA and EPS09 predictions.

Photoproduction of $J/\psi$ and $\psi(2S)$ vector mesons on nuclei can be studied 
in nucleus--nucleus ultraperipheral collisions (UPCs), 
when the colliding nuclei pass each other at large
impact parameters $|\vec{b}|$ so that the strong interaction between the nuclei is suppressed and 
they interact through the emission of quasi-real photons~\cite{Baltz:2007kq}.
The corresponding cross section reads:
\begin{eqnarray}
 \frac{d \sigma_{AA\to AA J/\psi [\psi(2S)]}(y)}{dy} 
=N_{\gamma/A}(y)\sigma_{\gamma A\to J/\psi [\psi(2S)] A}(y)+
N_{\gamma/A}(-y)\sigma_{\gamma A\to J/\psi [\psi(2S)] A}(-y) \,,
\label{csupc}
\end{eqnarray}
where $N_{\gamma/A}(y)=\omega dN_{\gamma/A}(\omega)/d\omega$ is the photon flux; 
$y = \ln(2\omega/M_V)$ is the final vector meson rapidity, where $\omega$ is the photon 
energy.

The photon flux at large impact parameters $b=|{\vec b}| > 2R_A$
emitted by a fast-moving nucleus, $N_{\gamma /A}(\omega)$, 
can be approximated very well by the flux of equivalent 
photons produced by a point-like particle with the electric charge $Z$: 
\begin{equation}
N_{\gamma /A}(\omega)=\frac{2Z^2\alpha_{\rm e.m.}}{\pi} 
\int \limits_{2R_A}^{\infty} db\,
{X^2\over b} \left[K^2_1(X)+\frac {1} {\gamma^2_L} K^2_0(X)\right] \,,
\label{plflux}
\end{equation}
where $\alpha_{\rm e.m.}$ is the fine-structure constant; 
 $K_0(X)$ and $K_1(X)$ are modified Bessel functions;
$X=b \omega/\gamma_L$,  where $\gamma_L$ is the nucleus Lorentz factor.
The strong interactions between the
colliding nuclei is suppressed by the requirement that  $b>2R_A$.

\begin{figure}[t]
\centering
\epsfig{file=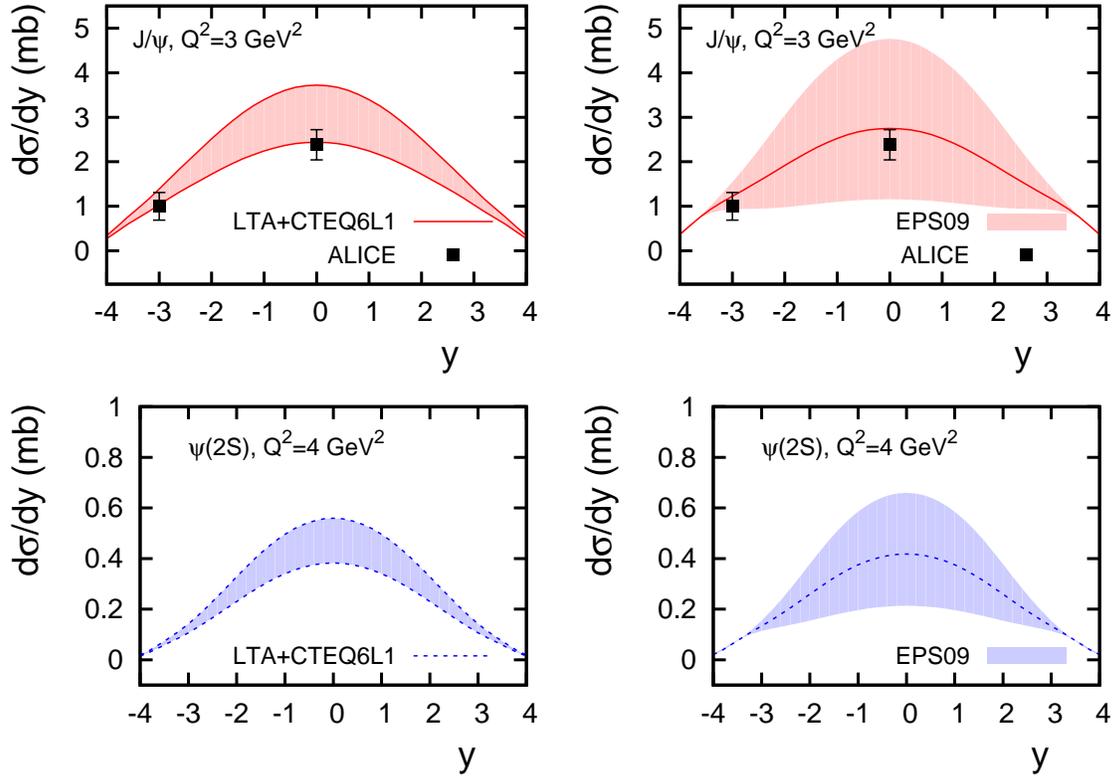,scale=1.5}
\caption{The cross sections of $J/\psi$ (upper panels) 
and $\psi(2S)$ (lower panels) 
photoproduction in Pb-Pb UPCs, $d \sigma_{AA\to AA J/\psi [\psi(2S)]}(y)/dy$, as a function of the 
vector meson rapidity $y$ at $\sqrt{s}=2.76$ TeV.
The filled squares with error bands are the ALICE data points~\cite{Abbas:2013oua,Abelev:2012ba}.
}
\label{fig:sigma_2s}
\end{figure}

Figure~\ref{fig:sigma_2s} presents our predictions for the cross sections of $J/\psi$ (two upper panels) 
and $\psi(2S)$ (two lower panels) 
photoproduction in Pb-Pb UPCs, $d \sigma_{AA\to AA J/\psi [\psi(2S)]}(y)/dy$, as a function of the 
final vector meson rapidity $y$ at $\sqrt{s}=2.76$ TeV. 
In the $J/\psi$ case, one can see that the theoretical predictions give the good description
of the ALICE data points~\cite{Abbas:2013oua,Abelev:2012ba} shown as filled squares with error bands.

Let us compare the $J/\psi$ and $\psi(2S)$ cross sections.
In particular, at the central rapidity $y=0$, we obtain (see Fig.~\ref{fig:sigma_2s}):
\begin{eqnarray}
&& \frac{d \sigma_{AA\to AA J/\psi}(y)}{dy}(y=0)=\left\{\begin{array}{ll}
2.44-3.72 \ {\rm mb} \,, & {\rm  LTA+CTEQ6L1} \\
2.75^{+2.01}_{-1.60} \ {\rm mb} \,, & {\rm EPS09} \,, 
\end{array} \right. \nonumber\\
&&\frac{d \sigma_{AA\to AA \psi(2S)}(y)}{dy}(y=0)=\left\{\begin{array}{ll}
0.38-0.56 \ {\rm mb} \,, & {\rm  LTA+CTEQ6L1} \\
0.42^{+0.24}_{-0.21} \ {\rm mb} \,, & {\rm EPS09} \,.
\end{array} \right.
\label{eq:sigma_y0}
\end{eqnarray} 
In the right-hand side of Eq.~(\ref{eq:sigma_y0}), we show the range of predicted values corresponding
to the intrinsic theoretical uncertainty: in the LTA+CTEQ6L1 case, the first value corresponds to the 
higher gluon shadowing scenario and the second one is for the lower gluon shadowing 
scenario~\cite{Frankfurt:2011cs}. In the EPS09 case, we show the central value with the associated
uncertainty due to the uncertainty of the extracted nuclear gluon PDF~\cite{eps09}.

Taking the ratio of the cross sections in Eq.~(\ref{eq:sigma_y0}), we obtain:
\begin{equation}
\frac{d \sigma_{AA\to AA \psi(2S)}(y)/dy(y=0)}{d \sigma_{AA\to AA J/\psi}(y)/dy(y=0)}
=\left\{\begin{array}{ll}
0.15-0.16 \,, & {\rm  LTA+CTEQ6L1} \\
0.15^{+0.03}_{-0.01} \,, & {\rm EPS09} \,.
\end{array} \right.
\label{eq:sigma_ratio_y0}
\end{equation}

These values can be be compared to: (i) the ratio of the $\psi(2S)$ and $J/\psi$ photoproduction 
cross sections on the proton,
and (ii) the ratio of $\psi(2S)$ and $J/\psi$ photoproduction cross sections 
in Pb-Pb UPCs calculated in the impulse approximation.

For the case (i), the H1 analysis gives~\cite{Adloff:2002re}:
\begin{equation}
\frac{\sigma_{\gamma p\to \psi(2S) p}}{\sigma_{\gamma p\to J/\psi p}}=0.166 \pm 0.007 ({\rm stat.}) \pm 0.008 ({\rm sys.}) \pm 0.007 ({\rm BR})
\label{eq:sigma_ratio_proton_y0}
\end{equation}
on the interval $40 < W_{\gamma p} < 150$ GeV.
Using the measured slopes of the $t$ dependence of the $J/\psi$ and $\psi(2S)$ diffractive photoproduction
cross sections, $B_{\rm el}^{J/\psi}=(4.99 \pm 0.13 \pm 0.39)$ GeV$^{-2}$ and 
$B_{\rm el}^{\psi(2S)}=(4.31 \pm 0.57 \pm 0.46)$ GeV$^{-2}$~\cite{Adloff:2002re},
one can find the ratio of the corresponding differential cross sections at $t=0$ and at the 
values of $W_{\gamma p}$ corresponding to $y=0$:
\begin{equation}
\frac{d\sigma_{\gamma p\to \psi(2S) p}/dt(t=0)}{d\sigma_{\gamma p\to J/\psi p}/dt(t=0)}=
\frac{B_{\rm el}^{\psi(2S)}}{B_{\rm el}^{J/\psi}}\frac{\sigma_{\gamma p\to \psi(2S) p}}{\sigma_{\gamma p\to J/\psi p}}=
0.157 \pm 0.032 \,,
\label{eq:sigma_ratio_proton_y0_2} 
\end{equation}
where the experimental errors have been added in quadrature. In 
Eq.~(\ref{eq:sigma_ratio_proton_y0_2}) we took into account that at $y=0$, 
$W_{\gamma p}$ in the $\psi(2S)$ case is slightly higher than that in the $J/\psi$ case.

For the case (ii), we obtain:
\begin{equation}
\frac{d \sigma_{AA\to AA \psi(2S)}^{\rm IA}(y)/dy(y=0)}{d \sigma_{AA\to AA J/\psi}^{\rm IA}(y)/dy(y=0)}
=0.133 \,.
\label{eq:sigma_ratio_IA_y0}
\end{equation}

A comparison of the results in Eq.~(\ref{eq:sigma_ratio_y0}) with those of 
Eqs.~(\ref{eq:sigma_ratio_proton_y0_2}) and (\ref{eq:sigma_ratio_IA_y0}) shows that
(i) the nuclear suppression of $J/\psi$ and $\psi(2S)$ photoproduction in Pb-Pb UPCs at central 
rapidities due to the nuclear gluon shadowing is rather similar, and (ii) 
in comparison to the $J/\psi$ case,
the effect of the slightly smaller nuclear shadowing suppression in the $\psi(2S)$ case is compensated
by a somewhat smaller photon flux, which makes the   
$(d \sigma_{AA\to AA \psi(2S)}(y)/dy(y=0))/(d \sigma_{AA\to AA J/\psi}(y)/dy(y=0))$ ratio numerically close
to the $(d\sigma_{\gamma p\to \psi(2S) p}/dt(t=0))/(d\sigma_{\gamma p\to J/\psi p}/dt(t=0))$ 
ratio for the free proton.

Finally, integrating the cross sections in Fig.~\ref{fig:sigma_2s} over the rapidity $y$ in the 
interval $-4 \leq  y  \leq 4$, we obtain for 
the ratio of the integrated cross sections:
\begin{equation}
\frac{\sigma_{AA\to AA \psi(2S)}}{\sigma_{AA\to AA J/\psi}}
=\left\{\begin{array}{ll}
0.13-0.14 \,, & {\rm  LTA+CTEQ6L1} \\
0.14^{+0.01}_{-0.01} \,, & {\rm EPS09} \,.
\end{array} \right.
\label{eq:sigma_ratio_yint}
\end{equation}

Photoproduction of the $\psi(2S)$ vector meson on nuclei in Pb-Pb UPCs at the LHC was also considered
in~\cite{Adeluyi:2013tuu} using an approach rather similar to ours in spirit but different in 
implementation and in~\cite{Ducati:2013bya} using the color dipole framework.
In contrast to our results, it was found in~\cite{Adeluyi:2013tuu} 
that the ratio of the $\psi(2S)$ and $J/\psi$ photoproduction cross sections in Pb-Pb UPCs is noticeably
enhanced compared to the free nucleon case. 

In Ref.~\cite{Ducati:2013bya}, when shadowing suppression of the cross sections of 
$J/\psi$ and $\psi(2S)$ photoproduction in Pb-Pb UPCs is evaluated using the standard dipole formalism
based on multiple scattering of $q {\bar q}$ dipoles on target nucleons, the obtained suppression factor
is small in both cases and is significantly smaller than predicted in our approach, see the
discussion in~\cite{Guzey:2013xba,Guzey:2013qza}.
 As a result, the dipole formalism without gluon shadowing ($R_g=1$) fails to reproduce the ALICE $J/\psi$
UPC data~\cite{Abbas:2013oua,Abelev:2012ba}. To remedy this, an additional suppression factor of 
$R_g < 1$, which aims to take into account large gluon shadowing, was also included 
in the analysis~\cite{Ducati:2013bya},
which in our opinion constitutes possible double counting.
The resulting model predicts large and very similar suppression of $J/\psi$ and $\psi(2S)$ 
photoproduction in Pb-Pb UPCs at the LHC, which is in agreement with our findings.

In summary, within the leading logarithmic approximation of perturbative QCD 
coupled with the leading twist approach to nuclear shadowing,
we compared the cross sections of photoproduction of $J/\psi$ and $\psi(2S)$ mesons on nuclei 
in Pb-Pb UPCs in the LHC kinematics.
We found that the nuclear suppression of these cross sections
due to the nuclear gluon shadowing is large and rather similar 
in the $J/\psi$ and $\psi(2S)$ cases.


\begin{thebibliography}{99}

\bibitem{Guzey:2013xba}
  V.~Guzey, E.~Kryshen, M.~Strikman and M.~Zhalov,
  Phys.\ Lett.\ B {\bf 726} (2013) 290
  [arXiv:1305.1724 [hep-ph]].

\bibitem{Guzey:2013qza}
  V.~Guzey and M.~Zhalov,
  JHEP {\bf 1310} (2013) 207
  [arXiv:1307.4526 [hep-ph]].

\bibitem{Abbas:2013oua} 
  E.~Abbas {\it et al.}  [ALICE Collaboration],
  Eur.\ Phys.\ J.\ C {\bf 73} (2013) 2617 
  [arXiv:1305.1467 [nucl-ex]].

\bibitem{Abelev:2012ba} 
  B.~Abelev {\it et al.}  [ALICE Collaboration],
  Phys.\ Lett.\ B {\bf 718} (2013) 1273 
  [arXiv:1209.3715 [nucl-ex]].


\bibitem{Ryskin:1992ui}
  M.~G.~Ryskin,
  Z.\ Phys.\  {\bf C57} (1993) 89.

\bibitem{Ryskin:1995hz}
  M.~G.~Ryskin, R.~G.~Roberts, A.~D.~Martin and E.~M.~Levin,
  Z.\ Phys.\ C {\bf 76} (1997) 231
  [hep-ph/9511228].


\bibitem{Martin:2007sb}
  A.~D.~Martin, C.~Nockles, M.~G.~Ryskin and T.~Teubner,
  Phys.\ Lett.\ B {\bf 662} (2008) 252
  [arXiv:0709.4406 [hep-ph]].


\bibitem{Adloff:2002re}
  C.~Adloff {\it et al.}  [H1 Collaboration],
  Phys.\ Lett.\ B {\bf 541} (2002) 251
  [hep-ex/0205107].

\bibitem{Zulkaply:2012oaa}
  Z.~B.~Zulkaply,
  ``Exclusive photoproduction of $\psi$(2S) in electron-proton collision at HERA,''
Ph.D. thesis (2012).


\bibitem{Jones:2013eda}
  S.~P.~Jones, A.~D.~Martin, M.~G.~Ryskin and T.~Teubner,
  J.\ Phys.\ G {\bf 41} (2014) 055009
  [arXiv:1312.6795 [hep-ph]].

\bibitem{Pumplin:2002vw}
  J.~Pumplin, D.~R.~Stump, J.~Huston, H.~L.~Lai, P.~M.~Nadolsky and W.~K.~Tung,
  JHEP {\bf 0207} (2002) 012
  [hep-ph/0201195].


 \bibitem{Frankfurt:2011cs}
  L.~Frankfurt, V.~Guzey and M.~Strikman,
  Phys.\ Rept.\ {\bf 512} (2012) 255
  [arXiv:1106.2091 [hep-ph]].

 \bibitem{eps09}
 K.~J.~Eskola, H.~Paukkunen and C.~A.~Salgado,
  JHEP {\bf 0904} (2009) 065
  [arXiv:0902.4154 [hep-ph]].


\bibitem{Baltz:2007kq} 
  A.~J.~Baltz {\it et al.},
  Phys.\ Rept.\  {\bf 458} (2008) 1
  [arXiv:0706.3356 [nucl-ex]].

\bibitem{Adeluyi:2013tuu}
  A.~Adeluyi and T.~Nguyen,
  Phys.\ Rev.\ C {\bf 87} (2013) 027901
  [arXiv:1302.4288 [nucl-th]].

\bibitem{Ducati:2013bya}
  M.~B.~G.~Ducati, M.~T.~Griep and M.~V.~T.~Machado,
  Phys.\ Rev.\ C {\bf 88} (2013) 014910
  [arXiv:1305.2407 [hep-ph]].

 

\end{thebibliography}
\end{document}